\documentclass[onecolumn,english,natbib]{svjour3}
\usepackage{amsfonts}
\usepackage{amsmath}
\usepackage{amssymb}
\usepackage{graphicx}

\usepackage{subfig}
\begin{document}
\newcommand{\Zset}{\mathbb{Z}}
\newcommand{\BG}{\mathcal{N_B}}
\newcommand{\kv}{\mathbf{k}}
\newcommand{\rv}{\mathbf{r}}
\newcommand{\bv}{\mathbf{b}}

\titlerunning{Analysis of dislocations in quasicrystals} 

\title{Analysis of dislocations in quasicrystals composed of self-assembled
nanoparticles}

\author{Liron Korkidi \and Kobi Barkan \and Ron Lifshitz }

\institute{Raymond and Beverly Sackler School of Physics and Astronomy, Tel
Aviv University, Tel Aviv 69978, Israel}
\maketitle

\begin{abstract}
  We analyze transmission electron microscopy (TEM) images of
  self-assembled quasicrystals, composed of binary systems of
  nanoparticles. We use an automated procedure that identifies the
  positions of dislocations and determines their topological
  character. To achieve this we decompose the quasicrystal into its
  individual density modes, or Fourier components, and identify their
  topological winding numbers for every dislocation.  This procedure
  associates a Burgers function with each dislocation, from which we
  extract the components of the Burgers vector after choosing a
  basis. The Burgers vectors that we see in the experimental images
  are all of lowest order, containing only 0's and 1's as their
  components. We argue that the density of the different types of
  Burgers vectors depends on their energetic cost.
\end{abstract}

\section{Dislocations in self-assembled soft-matter quasicrystals}

Self-assembled soft-matter quasicrystals have been observed in recent
years in a wide variety of different systems, in all cases but one
with dodecagonal (12-fold) point-group symmetry.  First discovered
by~\cite{zeng04} in liquid crystals made of micelle-forming
dendrimers, self-assembled soft-matter quasicrystals have since
appeared in other systems such as ABC-star
polymers~\citep{hayashida07}, in binary systems of
nanoparticles~\citep{TalpainNature}, in block co-polymer
micelles~\citep{18fold}, and in mesoporous silica~\citep{xiao12}. These newly-realized
systems not only provide exciting platforms for the
fundamental study of the physics of quasicrystals
\citep{barkan11}, they also hold the promise for new and exciting
applications, especially in the field of photonics.  An overview of
soft matter quasicrystals, including many references relevant to these
systems, is given by \cite{LD07} as well as by Ungar et
al.~(\citeyear{ungar05}, \citeyear{ungar11}) and
\cite{Dotera11,Dotera12}.

Here we concentrate on the systems of nanoparticles studied by
\cite{TalpainNature}, consisting typically of two types of particles,
such as PbS, Au, Fe$_{2}$O$_{3}$, and Pd, with different
diameters. These binary systems of particles, when placed in solution,
self-assemble into structures with long-range order, including 12-fold
symmetric quasicrystals. The dimensions of the particles---typically a
few nanometers in diameter---are such that they can be imaged directly
using a transmission electron microscope (TEM). This allows one to
study effects that are inaccessible with atomic-scale
quasicrystals. Here we present a quantitative analysis of the
dislocations that are naturally formed in these quasicrystals as they
self-assemble.

In periodic crystals in $d$-dimensions one can usually identify the
position of a dislocation rather easily by the termination of a plane
of atoms in three dimensions, or a line of atoms in two
dimensions. One then chooses a basis for the periodic lattice;
encircles each dislocation with a Burgers loop, or a Burgers circuit,
of basis vectors; and counts the accumulated difference between the
number of steps taken forward and backward in the direction of each of
the $d$ basis vectors. The $d$ integers thus obtained define the
Burgers vector which encodes the topological character of the
dislocation. A similar real-space procedure can be used on a
quasiperiodic crystal by overlaying it with a quasiperiodic tiling of
rank $D>d$ (for a definition, see below), yielding a $D$-component
Burgers vector. A tiling-based analysis of binary systems of
nanoparticles was indeed recently carried out by
\cite{TalpainDislocations}. Here we propose an alternative approach
for analyzing dislocations in Fourier space that we believe is useful
when dealing with aperiodic crystals.

\section{Density modes, winding numbers, and the Burgers function}

Let us describe the density of nanoparticles in a self-assembled
crystal by a function $\rho(\rv)$. The Fourier expansion of such a
function is given by
\begin{equation}\label{eq:DensityModes}
\rho(\rv)=\underset{\kv\in L}{\sum}\rho(\kv)e^{i\kv\cdot\rv},
\end{equation}
where the (reciprocal) lattice $L$ is a finitely generated
$\Zset$-module, which means that it can be expressed as the set of all
integral linear combinations of a finite number $D$ of $d$-dimensional
wave vectors, $\bv^{(1)},\ldots,\bv^{(D)}$. In the special case where
the smallest possible $D$, called the \emph{rank} of the crystal, is
equal to the physical dimension $d$, the crystal is periodic. More
generally, for quasiperiodic crystals $D\geq d$, and we refer to all
quasiperiodic crystals that are not periodic as \emph{quasicrystals}
\citep[see][]{lifshitz03,lifshitz07}.

As explained elsewhere \citep{RonIsraelChem}, the topological nature
of a dislocation is related to the fact that it cannot be made to
disappear by local structural changes. For this to be the case, as one
follows a loop around the position of a dislocation and returns to the
point of origin, one sees a crystal that is everywhere only-slightly
distorted from the perfectly ordered state, except near the core of
the dislocation. In particular, the complex amplitudes $\rho(\kv)$ of
the density modes maintain their magnitudes along the loop, each
accumulating at most a phase, which upon return to the point of origin
must be an integer multiple of $2\pi$. The collection of all such
integers, or so-called \emph{winding numbers}, for a given dislocation
defines a linear function $\BG(\kv)$ from the lattice $L$ to the set
of integers $\Zset$, which we call the \emph{Burgers function}.

The Burgers function of a given dislocation associates a particular
winding number $\BG(\kv)$ with every wave vector $\kv\in L$. Because
this function is linear, after choosing a basis $\{\bv^{(i)}\}$ for
the lattice, it is uniquely specified by a set of only $D$ integers
$n_i\equiv \BG(\bv^{(i)})$, forming the \emph{Burgers vector}
$(n_1,\ldots,n_D)$. Thus,
\begin{equation}\label{eq:BG}
  \forall \kv = \sum_{i=1}^D a_{i} \bv^{(i)} \in L:\quad \BG(\kv) 
 =\sum_{i=1}^D a_{i} \BG(\bv^{(i)}) = \sum_{i=1}^D a_{i} n_i,
\end{equation}
where $a_i\in\Zset$. This implies that in order to fully characterize
a dislocation in an experimental image it suffices to isolate the $D$
density modes associated with a chosen basis, and obtain their
corresponding winding numbers. This is the basis of the approach
presented below for analyzing dislocations \citep[for more detail,
see][]{Gilad,Freedman06,Freedman07}.

\section{Analysis of the dislocations in a quasicrystal of nanoparticles}

\begin{figure}
\subfloat[]{\includegraphics[width=0.33\textwidth]{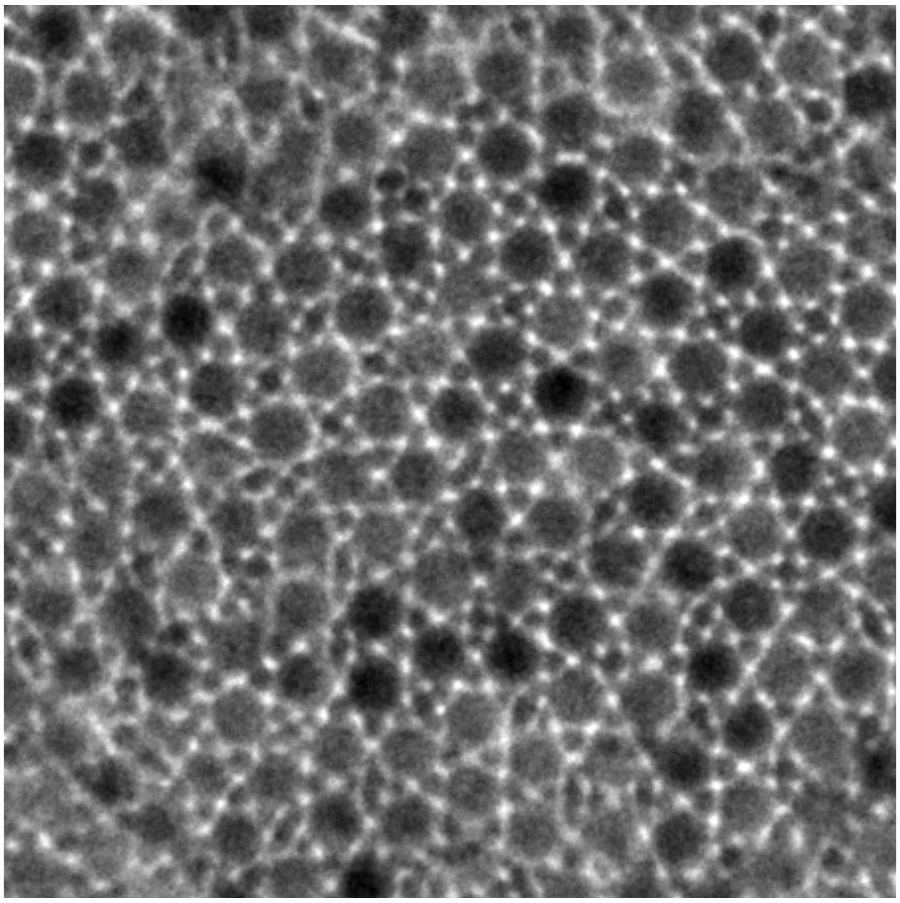}}
\subfloat[]{\includegraphics[width=0.33\textwidth]{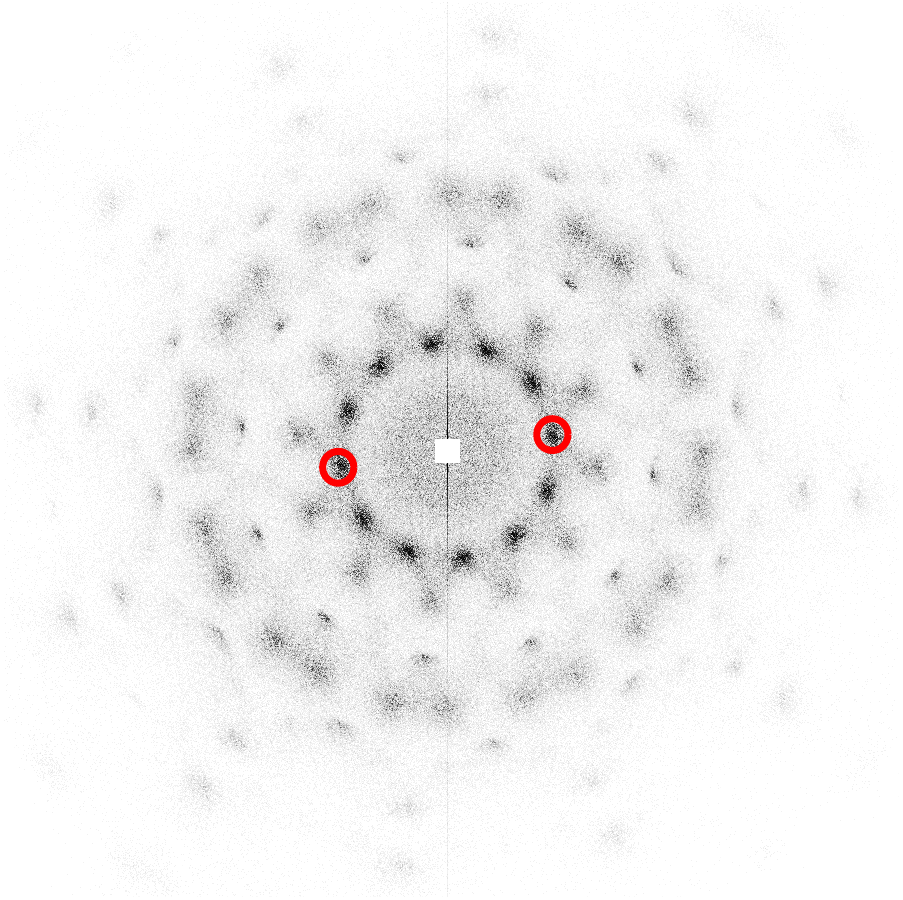}}
\subfloat[]{\includegraphics[width=0.33\textwidth]{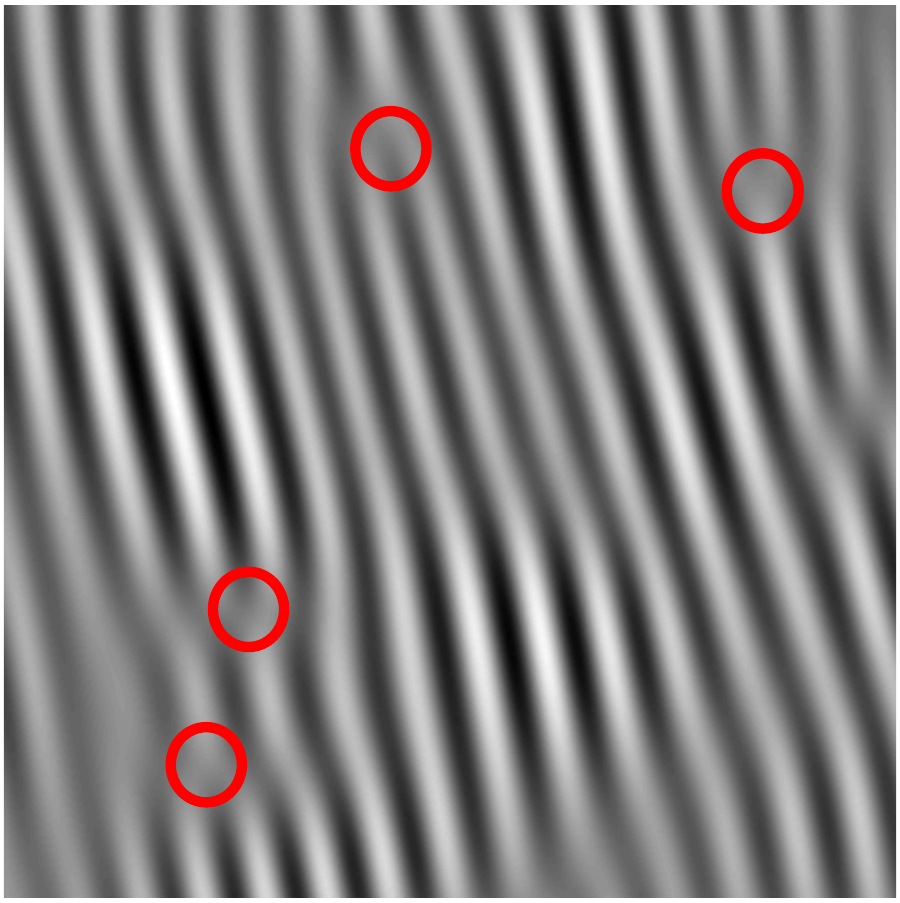}}
\caption{\label{fig:Procedure}(a) A section of a TEM image of a
  dodecagonal quasicrystal, self-assembled from 11.2nm PbS and
  5.2nm Au showing several dislocations \citep[courtesy of
  Dmitri][]{TalapinTEM}. (b) Fourier transform of the TEM image in
  (a), with the central peak blocked. A pair of Bragg peaks,
  associated with one of the basis vectors and its negative, is marked
  in red. (c) The corresponding section of the inverse Fourier
  transform of the Bragg peaks marked in (b) with red circles marking
  the positions of four dislocations.}
\end{figure}

\begin{figure}
\begin{centering}
\includegraphics[width=0.4\textwidth]{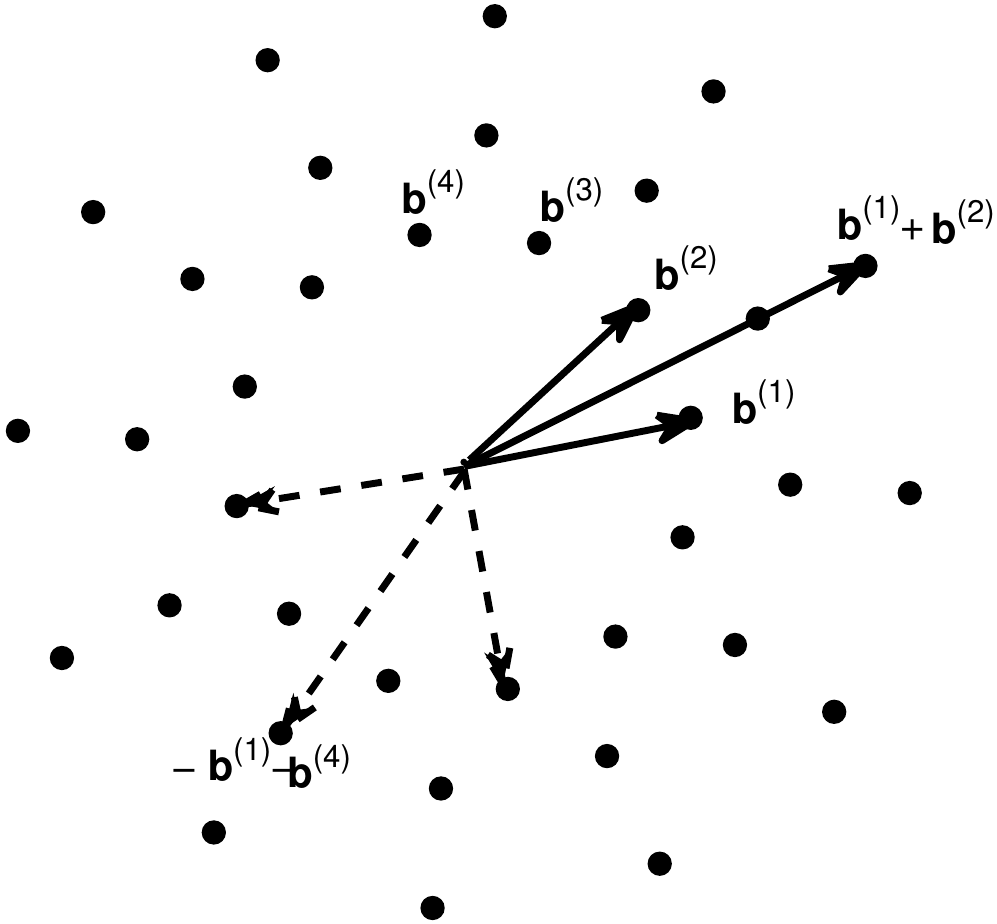}
\par\end{centering}

\centering{}\caption{\label{fig:The-basis-vectors} Schematic
  representation of the three strongest rings in the Fourier transform
  of our dodecagonal quasicrystal. The inner ring is the strongest,
  containing the four basis vectors $\bv^{(1)}\ldots\bv^{(4)}$.  The
  second strongest ring is the outer one, obtained from all the sums
  of two adjacent vectors in the inner ring, as indicated by solid
  arrows. The third strongest ring lies in between, obtained from all
  sums of vector pairs in the inner ring that are separated by 90
  degrees, as indicated by dashed arrows.}
\end{figure}

We begin with a high-resolution TEM image of one of the dodecagonal
quasicrystals grown by \cite{TalapinTEM}, a section of which is shown in
Fig.~\ref{fig:Procedure}(a). This particular quasicrystal is self-assembled from
11.2nm PbS and 5.2nm Au nanoparticles, and contains a distribution
of dislocations that are formed naturally during the self assembly. We
Fourier transform the TEM image, to obtain the diffraction image shown in
Fig.~\ref{fig:Procedure}(b), and then choose four of the Bragg peaks
in the 12-fold ring containing the strongest reflections as a
basis  $\bv^{(1)},\ldots,\bv^{(4)}$ for the reciprocal lattice. These
are labeled in the schematic representation of the lattice in
Fig.~\ref{fig:The-basis-vectors}.  

For each of the four pairs of Bragg peaks, associated with the chosen
basis vectors and their negatives, we then carry out the following
procedure:
\begin{enumerate}

\item We filter out small regions around the two opposite Bragg peaks,
  as indicated by a pair of red circles in
  Fig.~\ref{fig:Procedure}(b) for the case of the density mode
  associated with the wave vectors $\pm\bv^{(1)}$.

\item We inverse Fourier transform the filtered regions resulting in a
  real-space image of a single density mode. Dislocations appear as
  discontinuities in the stripes. We use a routine that identifies all
  the discontinuities and marks their positions, as shown in
  Fig.~\ref{fig:Procedure}(c) for this density mode.

\item For each dislocation, a second routine then extracts the $i^{\rm
    th}$ component $n_i=\BG(\bv^{(i)})$ of the Burgers vector. This is
  done by enclosing a counter-clockwise loop around its position and
  calculating the accumulated phase in units of $2\pi$. Practically
  what we do is count the number of stripes crossed
  moving in the direction of the wave vector $\bv^{(i)}$ on one side
  of the dislocation, and subtract the number of stripes crossed moving
  against the direction of $\bv^{(i)}$ while returning on the other
  side, as demonstrated in Fig.~\ref{fig:dislocations}.

\end{enumerate}
Finally, we verify the correctness of the calculation by extracting
the values $\BG(\kv)$ for additional wave vectors $\kv$ and checking that
they satisfy the linearity requirement given by Eq.~\eqref{eq:BG}.

\section{Results and discussion}

\begin{figure}
\subfloat[]{\begin{centering}
\begin{minipage}[t]{0.48\textwidth}%
\begin{center}
\includegraphics[width=0.48\textwidth]{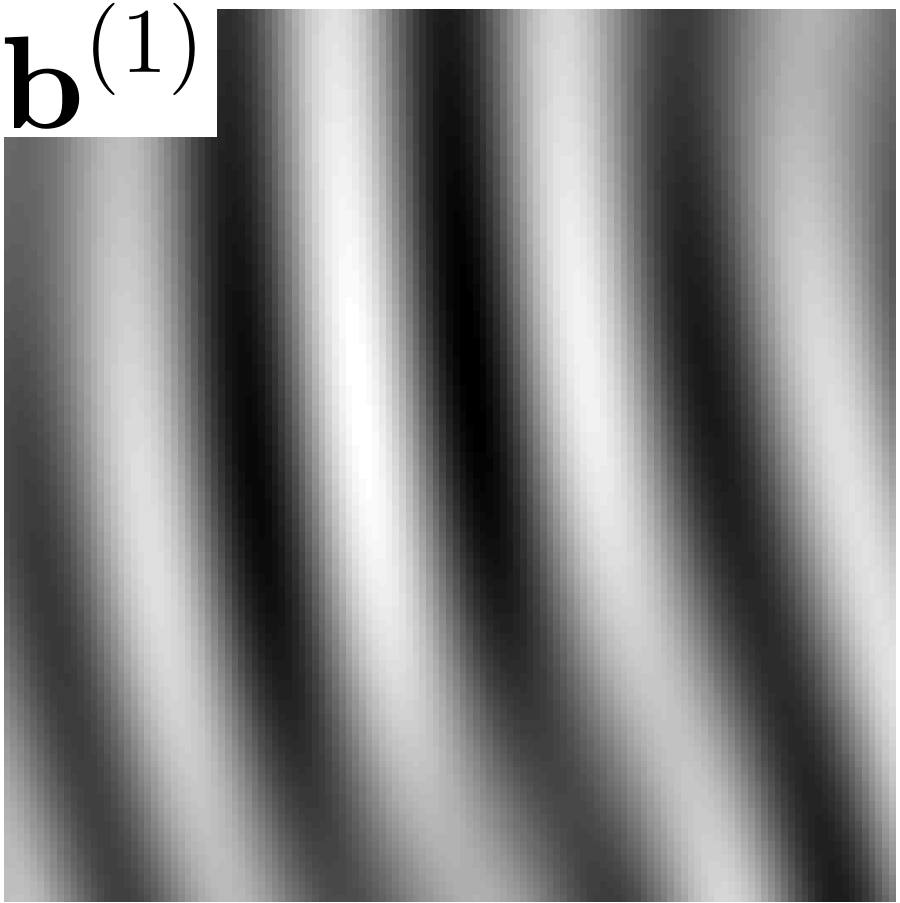}\hskip4pt%
\includegraphics[width=0.48\textwidth]{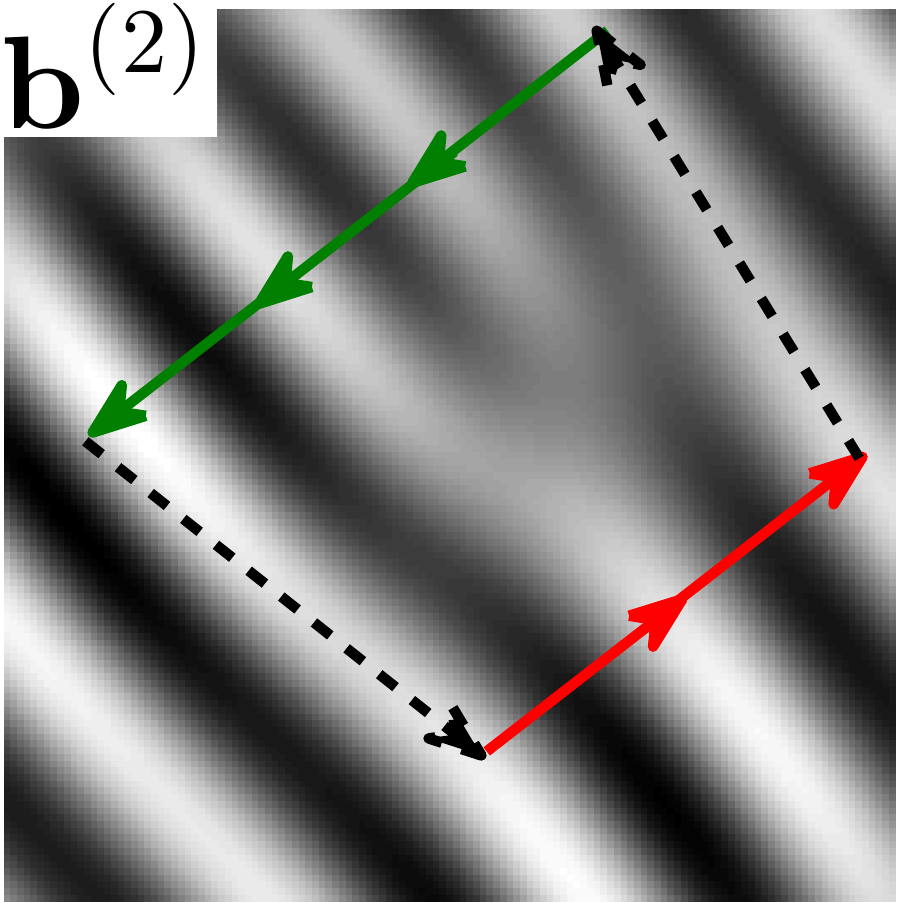}
\end{center}

\begin{center}
\includegraphics[width=0.48\textwidth]{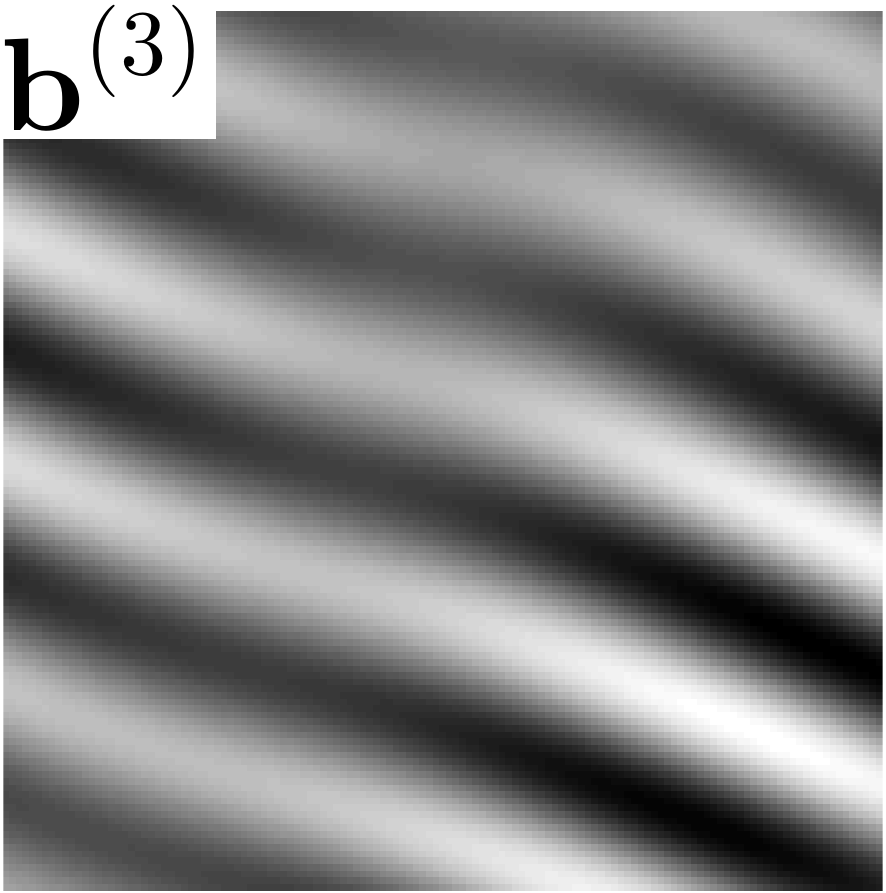}\hskip4pt%
\includegraphics[width=0.48\textwidth]{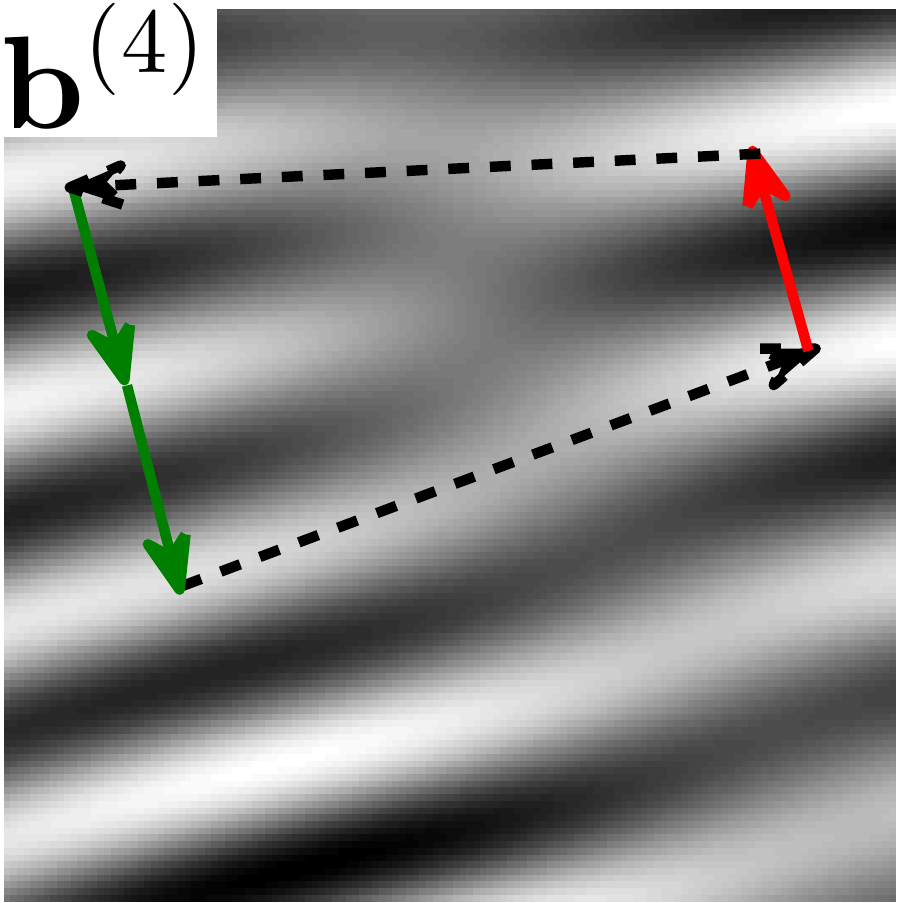}
\end{center}%
\end{minipage}
\end{centering}

\centering{}}\subfloat[]{\centering{}%
\begin{minipage}[t]{0.48\textwidth}%
\begin{center}
\includegraphics[width=0.48\textwidth]{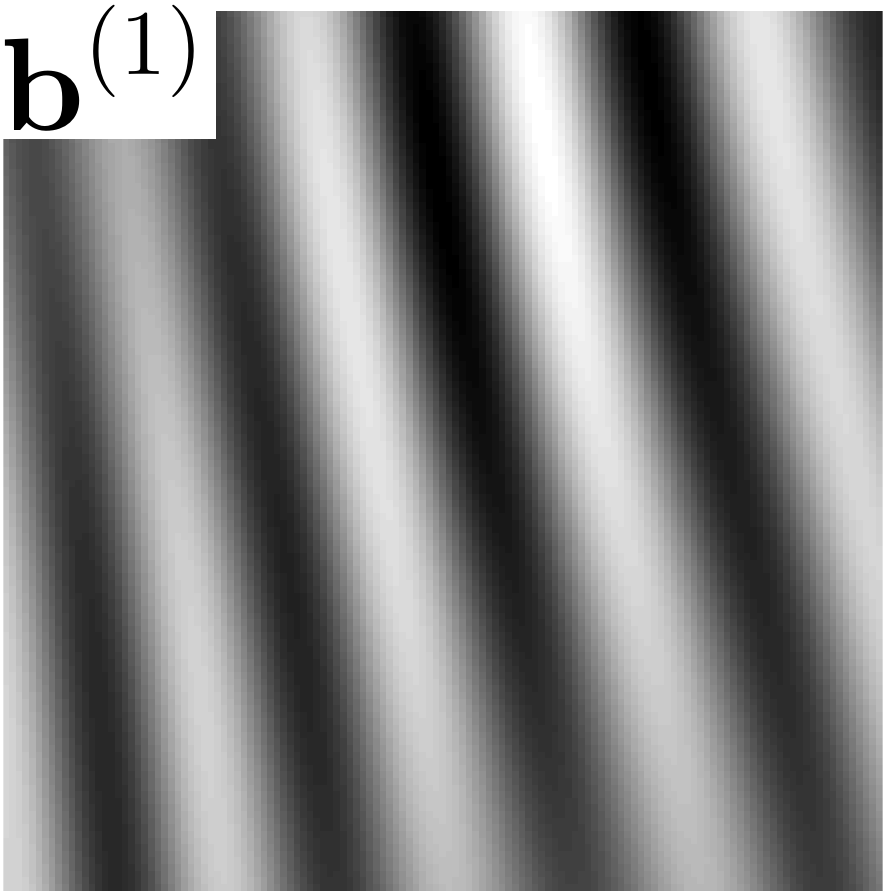}\hskip4pt%
\includegraphics[width=0.48\textwidth]{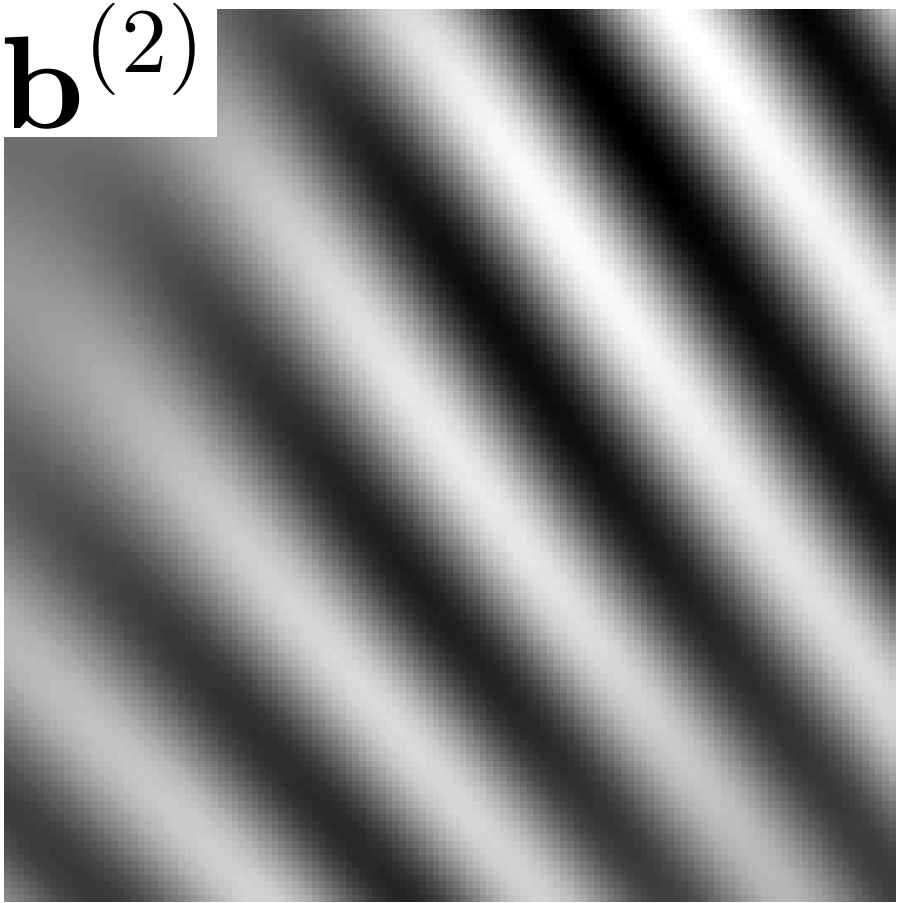}
\end{center}

\begin{center}
\includegraphics[width=0.48\textwidth]{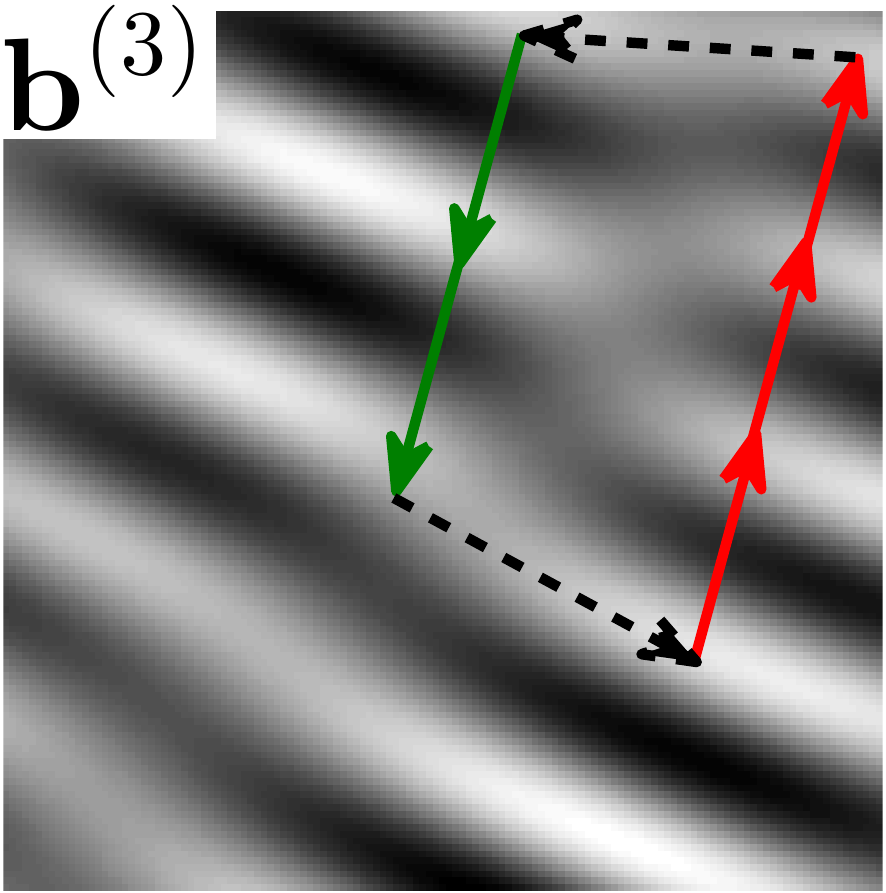}\hskip4pt%
\includegraphics[width=0.48\textwidth]{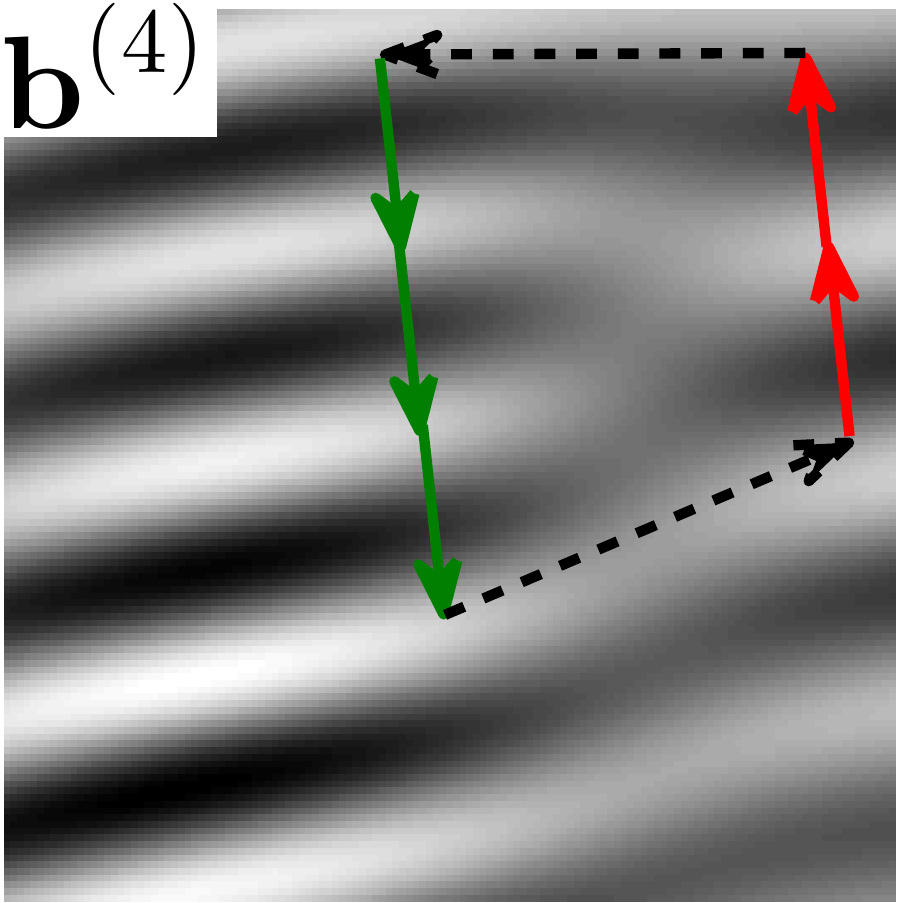}
\end{center}%
\end{minipage}}

\caption{\label{fig:dislocations} Two typical examples for the
  dislocations found using our procedure. In each counter-clockwise
  loop, the winding number $n_i$ is given by the number of red arrows,
  counting stripes crossed in the positive direction of $\bv^{(i)}$,
  minus the number of green arrows, counting stripes in the negative
  direction. (a)~A \textit{single-subset} dislocation with Burgers
  vector $\left(0,-1,0,-1\right)$ and (b) a \textit{dual-subset}
  dislocation with Burgers vector $(0,0,1,-1)$ (see text for
  definitions).}
\end{figure}

We typically find a density of a few dozen dislocations per $\mu
m^{2}$ in the nanoparticle quasicrystals of \citet{TalapinTEM}. All of
these dislocations are of lowest order in the sense that $n_{i} = 0$,
$1$ or $-1$. To understand the topological nature of these
dislocations it is useful to classify them by dividing the four basis
vectors into two hexagonal subsets---$\{\bv^{(1)},\bv^{(3)}\}$ and
$\{\bv^{(2)},\bv^{(4)}\}$ (see Fig.~\ref{fig:The-basis-vectors}). By
doing so we find that the density of dislocations with non-zero
components in only one of the subsets, which we call
\emph{single-subset} dislocations, is five times larger than that of
dislocations with non-zero components in both subsets, which we call
\emph{dual-subset} dislocations. Examples of the two types of
dislocations are shown in Fig.~\ref{fig:dislocations}.

To try and explain these observations, consider the free energy of the
self-assembled crystal as an expansion in products of density mode
amplitudes $\rho(\kv)$ \citep[see][]{RonIsraelChem},
\begin{equation}
  \label{eq:free}
  {\cal F}\{\rho\} = \sum_{n=2}^\infty \sum_{\kv_1\ldots\kv_n}
  A(\kv_1,\ldots\kv_n) \rho(\kv_1)\cdots\rho(\kv_n),
\end{equation}
where one can show that $A(\kv_1,\ldots\kv_n)=0$ unless $\kv_1 +
\ldots + \kv_n = 0$.  We argue that products in the sum
\eqref{eq:free} that contain high-intensity modes with large winding
numbers have a greater contribution to increasing the free energy away
from its minimum value in the perfect crystal. Accordingly,
high-intensity modes tend to exhibit smaller winding numbers. Indeed,
we find that all the winding numbers associated with the two brightest
rings (see Fig.~\ref{fig:Procedure}(b) and
Fig.~\ref{fig:The-basis-vectors}) are either 0 or $\pm1$, whereas it
is only on the 3$^{\rm rd}$ ring that we begin to see winding numbers
that are either 0, $\pm1$, or $\pm2$. Moreover, owing to the linearity
of the Burgers function, the fact that the winding numbers on the
second brightest ring are at most of magnitude 1 prevents two adjacent
peaks from the different subsets in the first ring from having
non-zero winding numbers of the same sign.  Because the ring of Bragg
peaks, obtained by adding pairs of wave vectors separated by 60
degrees, is extremely weak [see Fig.~\ref{fig:Procedure}(b)], there is
no such constraint on the winding numbers belonging to the same subset
of basis vectors. The fact that this constraint applies only to
dual-subset dislocations, reduces their possible combinations and
overall relative density.

\begin{figure}
\begin{center}
\subfloat[]{\includegraphics[width=0.4\textwidth]{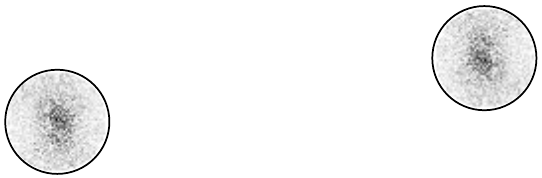}}\hskip2cm
\subfloat[]{\includegraphics[width=0.4\textwidth]{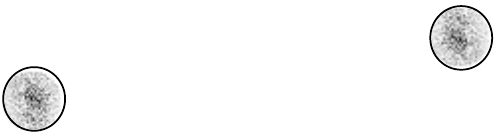}}\\
\subfloat[]{\includegraphics[width=0.4\textwidth]{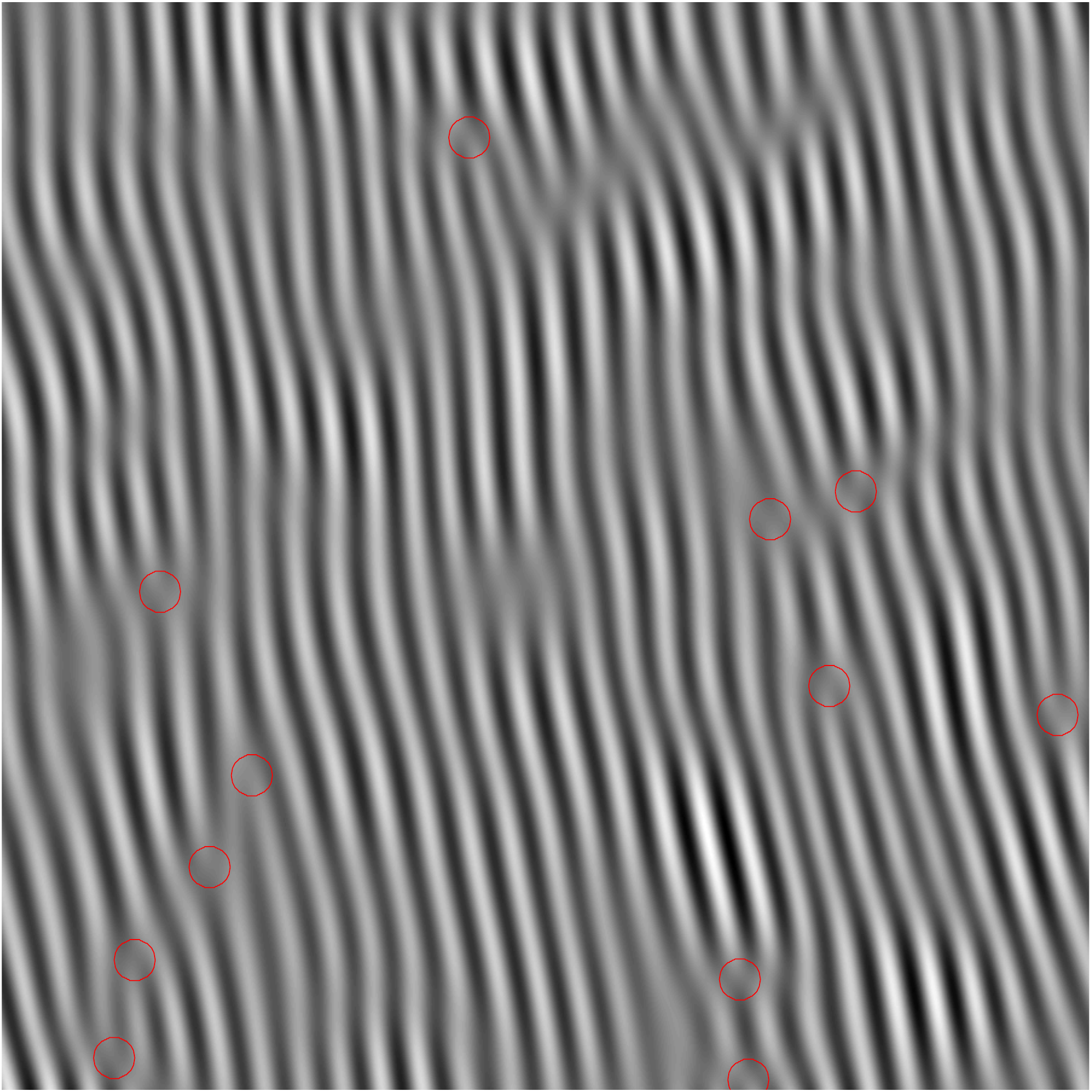}}\hskip2cm
\subfloat[]{\includegraphics[width=0.4\textwidth]{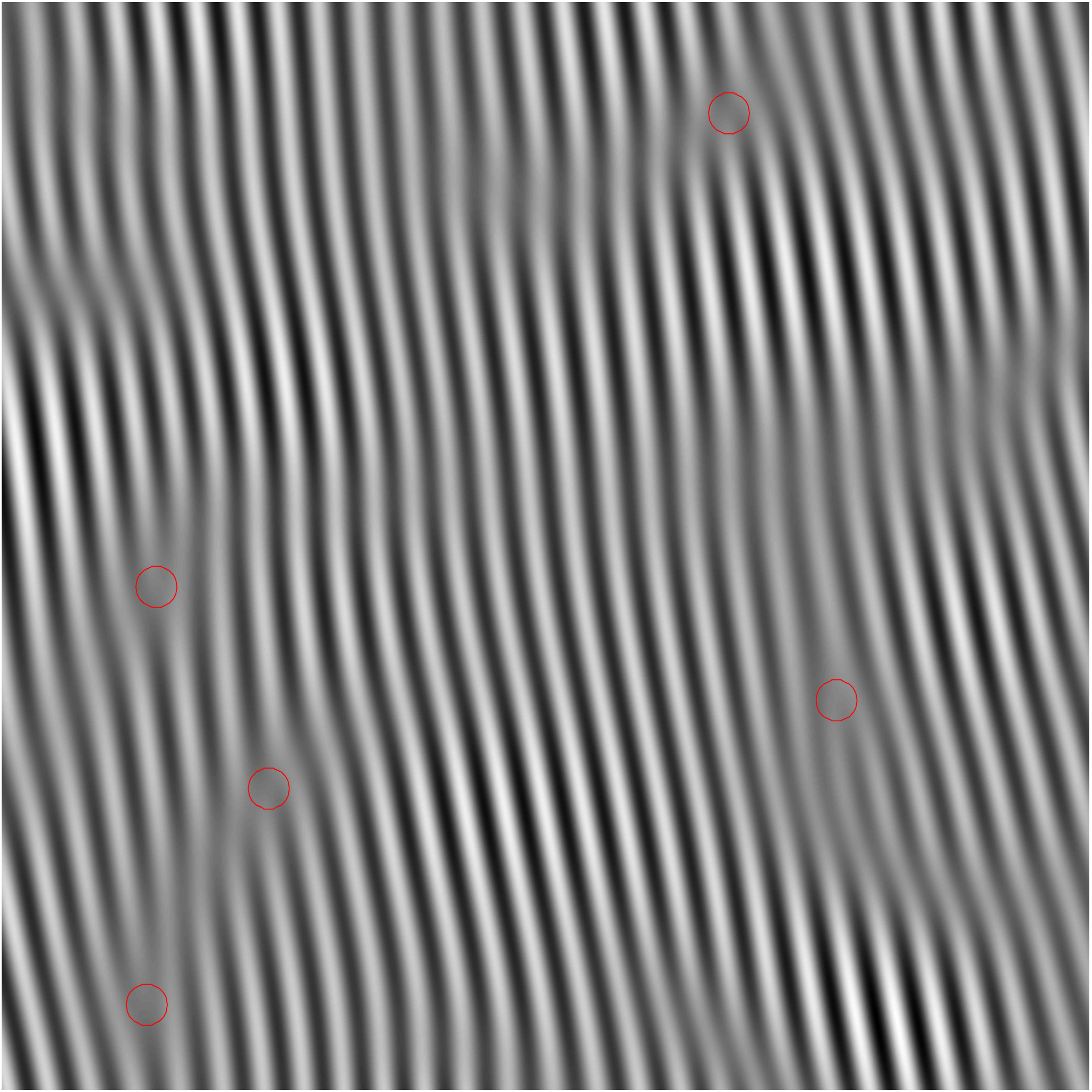}}
\end{center}
\caption{\label{fig:FilterSize} Inverse Fourier transform of a pair of
  Bragg peaks using filters of different size. For a circular filter
  with a radius of 25 pixels, used in (a), we find 12 dislocations,
  marked with red circles in (c). As we decrease the filter radius to
  15 pixels, in (b), we find only 5 dislocations in the inverse
  Fourier transform in (d).}
\end{figure}

A word of caution is in order regarding our approach for analyzing
dislocations. Because the density of the dislocations is relatively
high the Bragg peaks are not point-like but are rather spread as can
be seen in Fig.~\ref{fig:Procedure}(b).  This means that some of the
information about the dislocations may lie between Bragg peaks and
may be lost if the filters are too small.  Therefore, our approach is
sensitive to the shape and size of the filter that we use around each
Bragg peak. As we increase the filter size we obtain more information
and potentially find more dislocations, as demonstrated
in Fig.~\ref{fig:FilterSize}. We thus try to optimize the filter by
gradually enlarging its size until the number of dislocations stops
increasing substantially.

Our approach for analyzing dislocations should be easily adapted to
other systems even when the density of the dislocations is quite
large, as one may expect for soft matter systems. Moreover, for
dynamical systems that can be imaged in real time one can use our
automated method to follow and quantitatively analyze the dynamics of
the dislocations.

\begin{acknowledgements}
We are very grateful to Dmitri Talapin for providing the TEM images.
This research is supported by the Israel Science Foundation through
grant No.~556/10.
\end{acknowledgements}

\bibliographystyle{spr-chicago}
\bibliography{LironBib}

\end{document}